# When heart beats differently in depression: a review of HRV measures

Milena Čukić[1,*], Danka Savić[2] and Julia Sidorova[3]

[1]Instituto de Tecnología del Conocimiento, Universidad Complutense de Madrid, Spain.

[2] Vinča Institute for Nuclear Physics, Laboratory of Theoretical and Condensed Matter Physics

020/2, University of Belgrade, Belgrade, Serbia.

[3] Bioinformatics Platform, Hospital Clínic, Barcelona, Spain.

*Corresponding author on behalf of other co-authors is Milena Čukić, MSc Eng and PhD, Orchid 0000-0002-9162-987X (Current Address: Koningin Wilhelminaplein 644, 1062 KS, Amsterdam, The Netherlands; Tel. +31615178926, email: micukic@ucm.es, micu@3ega.nl )

**Abstract**

Background and Objective: The connection between depression and autonomous nervous system (ANS) is well documented in scientific literature. Heart rate variability (HRV) is a rich source of information for studying the dynamics of this relation. Disturbed heart dynamics in depression seriously increases mortality risk. Technical sciences help improve early detection and monitoring and offer more accurate management of treatment. Based on advances in computational power, information theory, complex systems dynamics, and nonlinear analysis applied to physiological complexity, we can now turn to novel biomarkers extracted from electrophysiological signals. This work is a cross-sectional analysis with methodological commentary of application of nonlinear measures of HRV related to depression.

Methods: We systematically searched online for papers exploring the connection of depression and HRV, using both conventional and nonlinear analysis. We scrutinized chosen publications and methodologically analyzed and compared them.

Results: Sixty-seven publications on the connection of HRV measures and depression meeting our inclusion criteria are selected. The effectiveness of the applied methods of electrocardiogram (ECG) analysis are compared and discussed in the light of detection and prevention of depression-related cardiac diseases.

Conclusions: It is clear that aberrated ANS dynamics can be detected via ECG analysis, where nonlinear measures show to be more sensitive and accurate than standard time and spectral ones. With the portable ECG devices and cloud-based telehealth applications, monitoring of outpatients could be done anytime anywhere. We appeal for early screening for cardiovascular abnormalities in depressive patients to prevent possible deleterious events.

**Key words:** Heart rate variability (HRV), Electrocardiogram (ECG), Depression, Autonomous nervous system (ANS), Nonlinear measures, Cardiac risk

**Introduction**

Cardiovascular diseases (CVDs) are the number one cause of death globally according to World Health Organization (1). Depression is the number one of the mental health-related contributors to the global burden of disease (2). When combined, these two diseases can lead to increased mortality risk (3, 4, 5). Recently, European Society of Cardiology published a position paper about the mechanisms linking depression and coronary heart disease (CHD), based on abundant evidence from literature (6). Although this connection was discovered a long time ago (7, 8, 9), the screening of depression patients on CHD is still far from routine.

In nearly 70% of patients with depression, somatic symptoms such as lack of energy, sleep disturbance, lack of appetite, decreased sex drive, general pains, etc. dominate the clinical picture (10). These symptoms are due to autonomous nervous system (ANS) dysfunction. Heart rate variability (HRV), regulated by ANS, is a marker of CHD. The relation between HRV and depression has been well understood (6, 11, 12, 13, 14).

Medical professionals interested in detection of depression may be uninformed about the knowledge and methods that might offer them additional insights about their patient's condition, the knowledge coming from theoretical research - mathematical analysis, complex systems dynamic theory, and information theory. These methods can be used to extract information embedded in electrophysiological signals, represented as time series - electrocardiogram (ECG), electroencephalogram (EEG), electromyogram (EMG), etc. Current view of what electrophysiological signals can yield is quite obsolete, limited by reductionist-approach analysis performed by biomedical devices' software standardly used at clinics.

A complex system is a system composed of many parts that act synergistically in a nonlinear fashion (their actions are not simple sums of the actions of the parts). Physiological systems are complex. Contrary to linear systems, 'even simple nonlinear system violate the principles of proportionality and superposition' (15). Although composed of multiple subunits, complex systems cannot be understood by analyzing these components individually: the main reason why reductionist approach fails is that subunits interact, and due to that interaction a very unpredictable behavior can be generated which simply defy explanation that traditional linear models can offer (15). The proportionality here, also, does not hold: even small initial changes to such a system, can lead to dramatic and unpredictable changes (colloquially known as the 'butterfly effect'). A very small increase or decrease of a certain parameter can cause abrupt transition from one behavior to another, often utterly irregular or highly periodic (16). Despite classical physiological paradigms based on homeostasis, where neural control mechanisms are designated to filter out noise and settle down in a stationary state, 'healthy heartbeat displays highly complex, apparently unpredictable fluctuations even under steady-state conditions' while heart failure, for example, shows 'slow periodic oscillations that correlate with Cheyne-Stokes breathing' (15). Theory of complex dynamics systems, widely known as the chaos theory, applies to such a system. Its behavior can be predicted at best for short intervals, although it is characterized with long-range correlations and ordered variability in healthy heart dynamics.

Standard methods of electrical signal analysis are designed for (predictable) mechanical systems. They rely on the assumption that the dynamics of the system may be linearized ("the noise" is cut off), in which process (in case of electrophysiological signals) a valuable information is lost. In information theory, the rate at which a system is producing information is described with Shannon entropy – a quantity reflecting the number of possible states the system can occur in, i.e. the level of uncertainty (unpredictability). Translated to signals, the higher the entropy, the higher the irregularity of a signal (more randomness). Unfortunately, a standardized definition of physiological complexity does not exist and most often it is taken that an irregular series has higher complexity. Thus, in general, entropy is a measure of signal irregularity that can be interpreted as complexity, and randomness (or pseudo randomness). This "awkward fact", as Vargas et al (16) note, is paradoxical as complexity assumes a structure which is the opposite of randomness. Early researchers of physiological complexity devised the test of the presence of chaos in electrophysiological signals (related to sequentonality), so it could be clearly identified as different from random noise (17). Nevertheless, as much as this confusion makes the insights into control mechanisms more difficult, the measures of complexity/irregularity differ between health and disease rendering them suitable for nonspecific markers of ill health.

Glass & Mackey (1988) stated that 'Random outputs result from degraded control mechanisms and/or breakdown of the coupling among them' (18). In one of the most extensive reviews of nonlinear measures extracted from electroencephalogram (EEG) to detect depression, De la Torre-Luque and Bornas (2017) concluded that the complexity of the signal in depression seems to be higher, but the system is probably more random than the healthy one (19). Neural control mechanisms, which demonstrate fractal properties, generate 'organized variability' (previously thought to be 'the noise' in the signal), characteristic for healthy physiological system (17). Physiological systems are scale-free; a self-similar fluctuations are observed on different time scales. From one moment to the next the fluctuations detected in the same signal are quite variable (15). A system which is fractal demonstrates irregularity across a wide range of scales; but that type of 'disorder' or 'roughness' on different scales is statistically similar. 'Organized variability is an inevitable consequence of fractal self-organization' as Goldberger described it (15). Another possibility is that higher EEG complexity is a consequence of a mechanism probably compensating for a deeper structure malfunction (20). In interpreting the results, the physical meaning of the applied nonlinear analysis and the physiological context of a particular disease have to be kept in mind. To stress again, nonlinear measures can serve as markers of the disorder regardless of the interpretation of the underlying mechanisms.

There is a whole field of well documented research on fractal and nonlinear analysis of heart rhythms (15, 17, 21-26). Cardiology is one of the rare disciplines that promptly embraced nonlinear analysis as an important tool for early and accurate diagnostics. Fractal and entropy measures showed to be very effective in detecting the slightest differences between healthy and ill heart dynamics, even when time series of compared states (healthy and for example, congestive heart failure ECG) are variating around almost the same mean values; indeed, they have tremendously different dynamics (15). Another impressive example is a case of detecting sudden infant death syndrome (SIDS) based on entropy measures calculated from ECG; standard method was not able to detect any difference between healthy and babies under a serious risk (25). Stephen M. Pincus, one of the first researchers to draw attention to the importance of understanding and interpretation of physiological complexity (22), adapted the Shannon entropy for cardiology research and devised the approximate entropy algorithm (AppEn), a statistic quantifying serial irregularity (25). Further, Richman and Moorman (2000) refined this measure into sample entropy (SampEn) (28), which was later improved by Costa (2002), proposing the multiscale entropy (MSE) algorithm, that calculates irregularity changes on multiple scales (physiological signals are 'scale-free') (29). Costa et al. (2006) performed a series of studies focusing on methods of analysis of ECG, and their work was a significant step in acceptance of nonlinear methodology (27). Can the successes of this innovation in cardiology be achieved in diagnosing depression? We believe that the answer is 'yes', because there already are many encouraging results on detecting the depression by nonlinear analysis of EEG (30, 31, 41).

Although computational neuroscience, systems biophysics, and some other computationally oriented disciplines exist (even computational psychiatry), the practitioners of clinical psychiatry, psychology or neurology are usually lacking the mathematical and physical background necessary for comprehension of these novel methods. It might be the reason why they are notoriously cautious to apply new methods, as happens, for example, with the very slow acceptance of machine learning, with engineers trying to speed up the acceptance process. Moreover, although standard spectral analyses are not well suited for (complex) physiological systems, as explained above, they are deeply established in clinical practice because most devices for recording physiological signals have a built-in algorithms based on Fourier analysis (32).

We made an overview (a cross-section analysis) of the most important studies that used the tools of various signal analyses to confirm the connection between HRV (as a marker of CHD) and depression. The papers with either classical or nonlinear or both analyses were included for comparison. The use of advanced nonlinear measures can be justified by demonstration of their capacity for early detection of cardiac risk in depression, with high sensitivity and accuracy. The goal of this work is to help convince clinicians to benefit from the mentioned methods (particularly nonlinear dynamics analyses), by including them in diagnostics and follow-ups to make these more precise and timely.

The nonlinear measures of HRV (such as DFA $\alpha_1$, Poincaré, SampEn) are confirmed to be reliable risk factors.

**Method**

Our search, performed in November 2020, included publications since the 1990s, when the term 'physiological complexity' was introduced by Peng, Pincus, Hausdorf, Goldberger and others. The intention was to search for applications of complexity measures in depression, potentially useful in clinical practice. Our query was kept as broad as possible in order to retrieve as many interesting papers as possible. We searched Google Scholar with logic formula: ('Depression' OR 'MDD' OR 'Bipolar depression' OR 'BDD') AND ('Electrocardiography' OR 'ECG' OR 'HRV' OR 'Heart rate variability') AND ('Nonlinear measures' OR 'Fractal dimension' OR 'Entropy').

We then scrutinized the abstracts and full texts (in English) and discarded irrelevant ones on the basis of our exclusion criteria which were: statistical analysis of medical history of the patients without analysis of electrophysiological data; other mood disorders where depression was not the main one; purely theoretical considerations of connection of HRV and depression without quantification of HRV measures; studies done on subjects under the age of 18; without age-matched controls; without peer review. Depression was diagnosed using DSM-III-R, DSM-IV-TR, DSM-5, MINI, BDI or MADRS. Some studies, found as references in hand searching of citation lists in review papers, were downloaded from Research Gate platform or were included even if nonretrievable as full text. Papers with standard analyses of ECG, added for comparison, were found in part via our online search, in part from reference lists of review papers, and some of them in full text from the authors who were kind to send us full texts.

We organized the results in a table with the basic data extracted from the included papers (the first author's name and the year of publication, the method and/or measures used in the research, the effect detected). We noted in the table whether a study detected significant effect or not.

After extensive cross-sectional comparison and interpretation of prior findings, we recommended practices in further work with depressive patients.

**Results**

During the search process, we first detected 7660 papers, 107 of which were selected for further screening, after reading the titles and abstracts. After reading the entire texts, we discarded another 40. Direct quantitative comparisons could not have been made, as studies varied in methodologies, as well as in research questions - detecting biomarkers or predictors of depression, effects of different therapies, etc. Therapies included those exploring medication effects/spillover (some studies compared two or more antidepressants' effects on HRV), psychological/psychiatric interventions, but also electromagnetic stimulation (electroconvulsive therapy/ECT, transcranial direct current stimulation/tDCS, repetitive transcranial magnetic stimulation/ rTMS, vagus nerve stimulation/VNS), and HRV biofeedback. As the effects of therapies are not the topic of this paper, we compared the studies by family of measures employed (standard-time and frequency/spectral domain, i.e. the classical Fourier approach-related measures vs fractal and nonlinear measures of HRV) and summarized their results/conclusions concerning only the detection of the depression-CVD (ANS) link. The studies encompassed those that use just time and/or frequency measures of HRV (35 papers), those that use both standard (time and frequency domain) and nonlinear measures (20 papers) and those that relied on nonlinear analysis only (12 papers).

**Table 1**. The summary of included publications (in chronological order) with information about the measures used and effect detected in every study. Standard- or conventional- measures of HRV are time and Frequency domain measures (the latter also known as spectral, typically employing Fourier spectral analysis) are numerous. In case that researchers used the whole battery of measures we simply implicated Time and Frequency domain measures of HRV, but mention them directly if researchers opted for just two or three specific measures in their work. Fractal and Nonlinear analysis cover entire families of measures, therefore we specified the names of specifically used measures in the cited research in the table. There are three groups of research included in our review: those relying just on conventional (time and frequency domain) measures, those that used exclusively nonlinear measures, and those that used both conventional and nonlinear measures comparing their effectivity of detection of changes in HRV.

Abbreviations:

*AIF*-Autonomic Information Flow

*AppEn, ApEn* - Approximate Entropy;

*CAD*-Coronary Artery Disease;

*CCM*-Complex Correlation Measure;

*CrossApEn* – Cross-approximate Entropy;

*CVC*-Cardiac Vagal Control;

*CVD*-Cardio Vascular Diseases;

*DFA*-Detrended Fluctuation Analysis;

*ECT* - Electroconvulsive Therapy;

*FD*-Fractal Dimension;

*GAD*-Generalized anxiety disorder;

*HF* - High Frequency;

*HR*-Heart Rate;

*HRV* – Heart Rate Variability;

*LF* - Low Frequency,

*LLE*-Largest Lyapunov Exponent

*MCD*-Mean Consecutive Difference;

*MCDSD*-standard deviation of mean consecutive difference;

*MDD*-Major Depressive Disorder;

*MEANNN* –average value of the NN intervals;

*MED*-Minimum Embedded Dimension

*ML*-Machine Learning.

*mRR-mean* R-R interval;

*MSE*-Multiscale Entropy;

*MSSD*- mean squared successive difference of R–R intervals;

*PCSD1*- the standard deviation of the Poincare plot perpendicular to the line of identity;

*pNN50*- the mean number of times an hour in which the change in successive normal sinus (NN) intervals exceeds 50ms;

*RCMSE*- Refined composite multiscale entropy of scale factor 1, 2, and 3;

*RMSSD* - root mean square of the successive differences;

*RQA*-Recurrence Plot Analysis;

*RSA*-Respiratory Sinus Arrhythmia;

*rTMS* - repetitive Transcranial Magnetic Stimulation;

*SampEn*-Sample Entropy;

*SDANN* - standard deviation of sequential five-minute R-R interval means;

*SDNN* –SD of the NN intervals;

*tDCS* - transcranial Direct Current Stimulation;

*TP* – Total Power;

*ULF* – Ultra Low Frequency;

*VHRB*-Variability Heart Rate Feedback;

*VLF* – Very Low Frequency;

*VNS*-Vagus Nerve Stimulation.

| Study | Analytical method | Conclusions/Findings (in depressed vs control, if not noted otherwise) |
|---|---|---|
| Dalack et al., 1990. (59) | Time and frequency-domain measures of HRV | Lower CVC |
| Yeragani et al., 1991. (60) | HR, MCD, MCDSD and PNN50 | No significant differences in standing resting state HR |
| Rechlin et al., 1994 (61) | Frequency domain measures | Higher HR and lower HRV |
| Lehofer et al., 1997. (62) | HR and HF | Higher HR; more heterogeneous vagal tone in melancholic type of MDD |
| Schultz et al., 1997. (63) | Frequency domain measures of HRV | Lower HRV in severe cases (increases after ECT) |
| Mooser et al., 1998. (64) | HR, HF, logRSA | More heterogeneous vagal tone; HR significantly greater, in MDD melancholic patients |
| Carney et al., 2000. (65) | Time and frequency-domain measures of HRV | Average heart rate and daytime rMSSD improved significantly in the severely depressed patients after CBT |
| Agelink et al., 2001. (66) | Time and frequency-domain measures of HRV | Of all tests a significant reduction in HRV could only be shown for the Valsalva ratio amongst the depressives compared to controls. |
| Yeragani et al., 2002 (67) | Frequency and nonlinear components of HRV, minimum embedded dimension (MED) Lyapunov exponents, Time-delay embedding and Attractor reconstruction | When only spectral powers of HP after treatment were used, total and VLF power contributed to the discrimination between the two treatments; when all nonlinear and spectral variables are entered, only 12-epoch SnetGS and 6-epoch average awake LLE were the significant discriminators. Nonlinear assessment may provide additional information about vagal modulation. |
| O'Connor et al., 2002. (68) | HR, RSA | Higher HR, lower HRV |
| Yeragani et al., 2002a. (69) | Frequency domain measures, fractal dimension (Katz) and symbolic dynamics | Decreased HF (increased after treatments) |
| Vigo et al., 2004. (70) | Time and frequency-domain measures of HRV, DFA, Poincaré plots and ApEn | $\alpha_1$ was higher, whereas SD1 and ApEn were lower in depression; $\alpha_1$ increased, while SD1 and ApEn decreased with worsening of the symptoms. Existing CAD confers a relative risk between 1.5 and 2.5 for cardiac morbidity and mortality |
| Bar et al., 2004. (71) | Time and frequency-domain measures of HRV | HR higher, RMSSD significantly lower |
| Pincus, 2006. (23) | AppEn and CrossAppEn | Both entropy measures are lower in BD |
| Rottenberg, 2007. (review) (72) | Time and frequency-domain measures of HRV review | The effect of depression on CVC is smaller than expected (only 2% of variance explained) |
| Udupa et al., 2007. (73) | Time and frequency-domain measures of HRV | rTMS improves HRV measures |
| Dawood et al., 2007. (74) | HR, LF, HF | HRV in unmedicated MDD patients indicated vagal function is not significantly impaired. |

| Siepmann et al., 2008. (75) | Time and frequency-domain measures of HRV | Biofeedback decreases HR, increases HRV |
|---|---|---|
| Van Zyl et al., 2008. (48) | Time and frequency-domain measures of HRV, review | Antidepressants decrease HR, increase HRV |
| Boetteger et al., 2008. (76) | Time and frequency domain measures and Autonomic Information Flow (AIF) | Analysis of standard HRV parameters in the time and frequency domain revealed no significant differences between groups. Area under the Autonomic Information Flow (AIF) curve ($INT^{NN}$) showed significant differences, in the morning hours only. |
| Licht et al., 2008. (77) | Time and Frequency-domain measures of HRV | Significantly lower HRV (due to antidepressants) |
| Baumert et al., 2009. (78) | Time and Frequency-domain measures of HRV, DFA, Symbolic measures, SampEn | No association found between NE spillover and standard HRV measures. Only complexity measures based on symbolic dynamics are correlated with the sympathetic outflow of the heart. Nonlinear metrics seem to reflect HRV features fundamentally different than standard ones. |
| Kikuchi et al., 2009. (79) | LF, HF, LF/HF | The power in all three sub-bands was lower in MDD than in HC. |
| Carney et al., 2009. review (80) | VLF, LnVLF, HR turbulence over 24h | Lower VLF; abnormal HR turbulence |
| Blasco-Lafarga, 2010. (81) | Time and frequency domain measures, DFA ($\alpha_1$) and SampEn | $\alpha_1$ had significant correlation with CCI score; measures that stratify fractal properties are more reliable than those that stratify HR complexity. |
| Taylor, 2010. (82) | Time and frequency domain measures, review | Higher heart rates and lower HRV |
| Kop et al., 2010. (43) | Time and Frequency domain measures, DFA ($\alpha_1$) and HR turbulence | Reduced $DFA_1$ and HR turbulence, but not standard measures. |
| Pizzi et al., 2010. (83) | Time-domain measures | SDANN and RMSSD significantly higher |
| Kemp et al., 2010. meta-analysis and review (13) | Time and frequency domain measures of HRV, AIF, LLE, DFA, Poincaré plots and ApEn | Depression without CVD is associated with reduced HRV, which decreases with increasing depression severity, most apparent with nonlinear measures of HRV. |
| Kemp, 2011a (35) | Time and frequency domain measures | Lower HRV |
| Schulz et al., 2010. (84) | Time and frequency domain measures, symbolic dynamics, compression entropy, MSE, DFA, Poincaré plot analysis | Time and frequency domain measures no significant differences; complexity indices from nonlinear dynamics demonstrated considerable changes in autonomous regulation |
| Jelinek et al., 2011. (85) | Poincare, Complex Correlation Measure (CCM) | CCM is more sensitive measure of HRV than Poincaré |
| Licht et al., 2011. (37) | Time and Frequency-domain measures of HRV | Lower HRV in depression |
| Kemp et al., 2011. (34) | Time and frequency domain measures | Lower HRV |

| Kemp et al., 2011a (35) | Time and Frequency-domain measures of HRV, Poincaré | Increased HR, lower HRV in depression suggesting disturbed dynamics. |
|---|---|---|
| Kemp et al., 2012. (44) | Time and Frequency-domain measures of HRV | Unmedicated, physically healthy MDD patients with and without comorbid anxiety had reduced HRV. Those with comorbid GAD showed the greatest reductions. |
| Migliorini et al., 2012. (86) | Time and frequency domain measures, DFA, Lempel-Ziv complexity 1/f slope and SampEn, ML | Scatter plots of SampEn and Lempel-Ziv values indicated separability in bipolar depression. Conventional measures did not make that distinction. |
| Valenza et al., 2012. (87) | Entropy measures | Decreased complexity resulting in pathological dynamics of ANS in depression. |
| Berger et al., 2012. (88) | Time and Frequency domain measures, RSA, and ApEn (RR and Resp), CrossApEn | Patients showed an increased heart rate and LF/HF-ratio. (RSA) and ApEnRR were reduced in patients in comparison to controls. |
| Chang et al., 2012. (89) | SampEn | MDD has distinctive psychophysiological profile |
| Brunnoni et al., 2013. (90) | RMSSD | Decreased HRV in MDD. |
| Moon et al., 2013. (91) | Time and Frequency domain measures of HRV and ApEn | ApEn was lower in depression than in controls. SDNN, RMSSD and HF were also lower in depression, but unable to differentiate between several psychiatric conditions. |
| Bozkurt et al., 2013. (92) | Frequency domain measures of HRV | Decreased HRV in very serious depression. |
| Tehardt et al., 2013. (93) | Frequency domain measures (excl. VLF) | Increased heart rate and reduced HRV in depression |
| Shinba, 2014. (94) | LF, HF, LF/HF | HR higher during the initial rest period; lesser response to the task; LF/HF was highest at rest |
| Kemp et al., 2014. (41) | Time and frequency domain measures and Poincaré plot, PSCD1 | MDD patients with melancholia (but not non-melancholia) displayed robust significant HR increase and lower resting-state HRV(RMSDNN, abs. HF power and SD Poincare) |
| Valenza et al., 2015. (95) | Instantaneous entropy measures of heartbeat dynamics | Reduced complexity in HRV |
| Chang et al., 2015. (46) | Spectral domain measures of HRV | Significantly lower mean R-R intervals, variance and LF, but higher LF/HF ratio in BDII |
| O' Regan et al., 2015. (95) | Time and frequency domain measures of HRV | No differences (nonmedicated depressed patients vs controls) |
| Lennartsson et al., 2016. (96) | Spectral domain measures of VHR | All measures, except LF/HF ratio were lower in clinical burnout. |
| De la Torre-Luque et al., 2016. review (97) | Frequency domain measures and varying nonlinear measures | Lower complexity/information and higher scale invariance; |
| Greco et al., 2016. (98) | Fractal dimension, SampEn, recurrence plot analysis (RQA) | Confirmed dysfunctional nonlinear sympato-vagal dynamics in mood disorders |
| Yeh et al., 2016. (38) | Frequency domain measures | Significantly lower HRV, LF and HF, higher LF/HF ratio |

| **Hage et al., 2017. (99)** | RSA and LF-HRV | Baseline heart period shorter (i.e., higher HR) in BDD |
|---|---|---|
| **Khandoker et al., 2017. (42)** | Multi-lag tone–entropy (T–E) analysis, regression tree | Mean entropy lower in MDD with suicide ideation |
| **Schumann et al., 2017. (100)** | Standard measures | Increased HR, no significant differences in HRV |
| **Valenza et al., 2017. (101)** | Lyapunov exponents, CV(IDLE) | Nonlinear IDLE measure is significantly increased in MDD, contrary to all standard measures. |
| **Kim et al., 2017. (102)** | Time and frequency domain measures of HRV | Low HRV, decreased HF, increased LF |
| **Chen et al., 2017. (103)** | Time and Frequency domain measures, Poincare plot/SD1, SD2 and RCMSE | RMSSD, SD1, SD2, and multiscale entropy (RCMSE3) lower in resting state and Valsalva test |
| **Shinba, 2017. (104)** | LF, HF, LF/HF | Low HF, high LF/HF at rest; sympathetic activation and reduced reactivity to task |
| **Carr et al., 2017 (105)** | mRR, SDNN, RMSSD, LF, HF, LF/HF | Positive mood in BD increases LF/HF, decreases HF power and mRR |
| **Caldwell & Steffen, 2018 (106)** | SDNN, LF, HF | A large increase in HRV after HRV biofeedback |
| **Borrione et al., 2018. (107)** | RMSSD, HF, LF, LF/HF | LF was predicted by component 4 of the HAM-D-17 (i.e. "core symptoms"), and LF/HF ratio by component 2 of the MADRS (i.e. "anhedonia/retardation") |
| **Byun et al., 2019. (108)** | SampEn, Lempel-Ziv complexity | Decreased measures of HRV and nonlinear non-redundant measures. |
| **Byun et al., 2019a. (109)** | ApEn, SampEn, FuzzyEn and Shannon, ML | All entropy measures significantly lower |
| **Saad M. et al., 2019. (39)** | Sleep derived pattern of heart-rate changes, ML | Detection of depression with accuracy of 79.9% |
| **Koch et al., 2019.** meta-analysis (33) | Time and frequency-domain measures, cross section analysis | All HRV-measures lower in depression. |

We opted to present the summary in chronological order to illustrate how the understanding of the link between depression and HRV measures evolved. It is interesting to note that some of the first studies reported to have calculated all the measures manually from paper ECG records, while the latest were typically using advanced nonlinear dynamical measures, advanced statistical techniques to correct for confounds, and machine learning to improve the accuracy of classification. Newer studies take (cardiac) mortality risk in depression as a well-known fact, even without referencing, while early ones just adopted measures like pNN50, as the latest innovation in cardiology. From the present body of literature, it is clear that the scientific community is gradually accepting the existence of the relation between depression and CVC. A debate about the role of antidepressant drugs in this dynamics lasted for several years, but the meta-analysis by Koch et al. (2019) of unmedicated samples (2250 patients and 1982 controls in total) clearly demonstrated reduced HRV in depressed patients without antidepressants (33). Regarding the controversies about unfavorable effects of antidepressants on HRV (13, 34, 35, 36, 37, 38, 39), this meta-analysis is making a cross-section analysis of what is known and accepted among researchers, and demonstrating an urge for translation of that knowledge to clinical practice.

**Discussion**

The set of papers presented here is only an illustration of the huge body of evidence of the relation between cardiovascular system dynamics and depression. The interconnection of coronary heart disease (CHD) and depression, probably dictated by ANS malfunction (decreased vagal activity and increased sympathetic arousal), greatly increases mortality risk that can be diminished by timely detection and adequate therapy of CHD in depressive patients. Screening CHD patients for depression is already included in clinical practice, why not also the other way around?

From the physiological complexity point of view, heart dynamics was the most studied in the previous decades and depression is becoming the most studied phenomenon at present. The introduction of complex systems dynamics and information theory concepts into this analysis sheds a new light on underlying problems and renders higher accuracies of detection by quantifying subtle ECG patterns and their changes. The data can be easily obtained by holter recordings or novel portable ECG monitoring devices that are approved as medical-grade signal quality equivalent to holter, but are much more practical and comfortable to use by the patient her-/himself, taking only a couple of minutes (e.g. Cardea3, Savvy, Gethemed, HealForce, Body metrics, etc.). The data can then be processed by a combination of nonlinear analytics and advanced statistical procedures (to control, for example, for comorbidities and other confounding factors, or for feature selection for further machine learning). Even better, the analysis can be empowered with machine learning applications that are widely in use due to high power of computation and cloud computing (for example, see 30 and 39).

Although there are many candidates for markers of cardiovascular diseases (including simple heart rate), most of the authors agree that a good indicator of CHD are heart rate variability (HRV) measures, and that HRV should be registered in patients diagnosed with clinical depression. Methodological examination of the papers shows that conventional (time and frequency domain) measures of the cardiological system dynamics are not as sensitive and reliable as the nonlinear ones. To mention just a few examples, Kemp et al. found that only nonlinear measures could discriminate between melancholic and non-melancholic depression (41); Schulz (84) showed that nonlinear dynamics considerably improved the detection of autonomic imbalances in comparison to standard linear indices of HRV that did not reach any statistically significant differences between MDD and healthy subjects; Khandoker et al. demonstrated that nonlinear (multi-lag tone–entropy) analysis of HRV was superior to linear analysis, predicting suicidal ideation with almost 95% accuracy when demographic factors were included (42). We can obtain another illustration from two review papers: for differences between depressed and healthy, Rottenberg (12) found a mild effect size in linear measures of HRV, while Kemp et al (13) found a large effect size for nonlinear measures of the same. Nevertheless, Koch et al. (2019), in their meta-analysis, found no evidence for a moderating role of study quality (33). This negative finding strengthens robustness of the observed HRV alterations in MDD (33).

Some of the studies in the table focused on depression treatment outcomes. The finding that treatments of depression by various antidepressants do not normalize HRV, despite alleviation of symptoms, made Kemp et al. (13) conclude their review with a warning message that there are remitted patients still at risk of CVD. Kop et al. (2010) observed the same in the Cardiovascular Health Study on 907 subjects (adding neuroimmunological markers to the research) and suggested a 'multifactorial approach for optimal identification and treatment of patients with high risk of cardiovascular mortality' (43). Given the controversies about 'unfavorable' effects of antidepressants on HRV (47), Kemp and colleagues (34) indicated several methodological mistakes and dissected with surgeon's precision the flaws of Licht's statistical inference: they showed how the inappropriate use of analysis of covariance applied in that study led to the conclusion that antidepressants are those that reduce HRV, while it could have been any of the covariates. Further, they confront the finding that all groups of antidepressants lead to 'unfavorable' effects on HRV with other researchers' results, e.g. van Zyl and colleagues' meta-analysis (48) showing that an increase in HRV and decrease in heart rate was associated with the use of SSRI, 'reflecting favorable effect on HRV'. Kemp et al. (13) and others (49, 50, 51), were also considering that 'different SSRI antidepressants might have different anticholinergic effects', and that SSRI might reduce CVD. The meta-analysis of Koch et al (33) clearly demonstrates that reductions in HRV are prevalent in unmedicated depressed patients without CVD.

Another advantage of HRV analysis in mental disorders might be in helping differential diagnosis or indicating comorbidities (44, 41). For example, Chang et al. (45) succeeded to distinguish between bipolar II depression and unipolar depression, based on SampEn analysis of HRV of 707 subjects. Bipolar patients are treated wrongly as having unipolar depression for eight years in average (46). Kemp et al. (34-36) found that anxiety disorders comorbid to MDD, most of all GAD, contributed to the reduction of HRV. They elaborated on how non-vagal components of heart-rate may further distinguish between disorder subtypes (41).

Cardiac vagal control (CVC) is associated with both physical and mental health. Low CVC is considered to be an indicator of risk of cardiac disease, including myocardial infarction and congestive heart failure (54). Rottenberg systematically explored the relation between CVC and depression, concluding that 'the use of CVC to predict the course of depression would lead to meaningful outcomes' (12). Since CVC reflects the extent of variability in heart rate that is gated by the respiratory cycle (12), it is logical to analyze its nonlinear dynamics and its aberrations, in order to detect and treat depression. It could be a link between Porges's Polyvagal Theory and Goldberger's Physiological Complexity (decomplexification & stereotypy of disease) approach to innovate and improve the detection and treatment of depression and to prevent associated CVD mortality risks. In parallel to polyvagal theory, there is also a neuroviscelar integration model (110) focusing on inhibitory cortico-subcortical circuit that enable adaptation required by the environmental changes, but both emphasize the importance of taking into account ANS aberrations, in line with existing need for improved psychiatric nosology.

Important insights about healthy heart dynamics, and how it changes with aging and disease were published in 90's, and served well the detection of several pathological entities (15, 55, 56). From this work we learn that mechanisms of neural control are fractal in nature (scale-free) and they generate the so-called organized variability (once believed to be a background noise to the signal), characteristic for healthy heart dynamics. In pathological states, a characteristic loss of complexity can be observed (decomplexification) (15) that leads to recognizable oscillatory (predictable) behavior of a complex system, reduced to a single scale or a frequency. The heart abnormalities can be seen as a loss of long-range correlations and an increase in regularity and predictability (15). This is what Goldberger refers to as the stereotypy of the disease, an easily recognized (diagnosed) characteristic behavior of an ill system, resulting in decreased ability to adapt to constantly changing environment. The aberrated dynamics can be precisely quantified by fractal and nonlinear measures because, as mentioned before, they are more adequate for accurate analysis of a nonlinear system than the conventional ones (57). The standard idea of comparison of healthy and ill organism pertains the calculation of traditional mean values, standard deviations and the like, from electrophysiological signal (here ECG). When one compares a healthy heart recording with the record of a patient diagnosed with congestive heart failure, their calculated values are within the same standard deviation from the mean value (like figure 2 in 15). Yet, it can be seen even with a naked eye that those signals are different; this difference can be best quantified with application of nonlinear measures. Traditional methods do not show a significant difference here. Pincus demonstrated the same inability of standard analysis to efficiently differentiate between aborted-SIDS infants and healthy ones (24). The stereotypy of disease, as Goldberger explained (15) is connected with the decomplexification of a dynamical system's output, observed in early complexity studies. Complexity analysis can complement this clinical heuristic with adequate mathematical tools to quantify the patient's state.

Another consequence of complexity (multiscale fractal-type variability) loss in disease is that the information content is degraded (15). Early papers on physiological complexity described this as well (23, 25, 26). Too aggressive preprocessing of the data can contribute to misleading results, again, aggravating the information loss from the signal. This is the reason why we advocate for analyzing segments of the records that are artifact-free, instead straining the data with additional artifacts introduced by artifact removal methods based on reductionist approach (broad-band signal is most information rich). The more complex the signal, the higher the entropy (implicating less of regularity in the system). In disease, therefore, we can use many entropy measures to detect whether a system is more or less regular - in many diseases the consequence of decomplexification is that the system becomes more predictable (regular, even periodic) and the output signal produced by such a system has those features too. In previous EEG-based detection of depression review, de la Torre- Luque speculated that the higher complexities detected with many nonlinear measures might implicate that due to changed intrinsic dynamics, the system's output actually exhibit the randomness that might be identified as higher complexity (19). The researchers that demonstrated high accuracy of detection of CVD risks used entropy measures, like sample entropy (SampEn), approximate entropy (AppEn), cross-entropy, multiscale entropy (MSE), generalized multiscale entropy (GMSE) and the like. SampEn actually calculates with very simple algorithm how predictable the signal is; in other words, how likely it is that two or three (or more) groups of samples would repeat again, in specified order and range of values in the original signal. Our results show that among many measures used in the reviewed literature, some stand out demonstrating easy classification between depressed and non-depressed individuals. Beside entropy measures, there are measures based on fractal analysis, like fractal dimension (Higuchi FD, or Katz FD), or detrended fluctuation analysis (DFA) that can identify long range fractal structure in the data. It is natural to expect that if the system (in this case system of neural control) exhibits fractal properties, the signal it produces has fractal properties too. If something is wrong with the control mechanisms, it must be reflected in the output signal; hence fractal analysis would be appropriate. Another efficient measure to detect mentioned aberrations is Poincaré plot analysis, which belongs to recurrence plot analysis family, able to quantify self-similarity in processes. Generalized Poincare analysis is also used with good results. Recurrence analysis is based partially on the idea that a complex dynamical system over the time revisits some states which are nearly, but never exactly identical states. When one applies this kind of analysis, a certain scatter plot is generated which can illustrate the variance of the data along the two most prominent axes, and we already know how to interpret that from many applications in physiology and cardiology. In the reviewed literature, there are also combinations and alterations of the mentioned methods of analysis, like combining the Poincare with DFA, or applying Pierson coefficients on prior Poincare analysis, or choosing the most prominent coefficients from several analyses and combining them as successful features for classification. Also, PCA analysis finds the most prominent components in the signal and it is shown that the first three PCAs can explain up to 95% of variance and increased number of components can explain up to 99% (this method is mainly used as exploratory data method in machine learning settings or for feature extraction to decrease the dimensionality of the problem).

It is important to mention another important property of electrophysiological signals, a historical order of the samples (meaning that one sample is affecting many consequent samples, hence their order is crucial information). Time series are therefore valuable source of information for analysis in physiology, and they exhibit statistical self-similarity. Researchers developed surrogate-data models for testing that, and also there are highly efficient methods to confirm the presence of chaos in the data (or nonlinearity) or even to find out whether the signal has a physiological origin or is simply a noise (stochastic signals do not possess intrinsic structure). Physiological signals reflect the intrinsic hierarchy of fractal structure of a system. Self-similar fractal structure dictates self- similarity in time series it produces. It became clear early in complexity studies that some previous moments in the evolution of the state of system dynamics affect the succeeding states; many of nonlinear measures mentioned here as the most successful (fractal measures, entropy measures, recurrence plots) exploit this for demonstrating the level of irregularity, complexity, clustering etc., because they take into account previous signal samples in the algorithms. This is probably the reason why spectral methods did not perform so efficiently in detecting the differences, because in any analysis based on Fourier analysis (which has reductionist approach), in the first step of calculation some of the components are lost (like demonstrated in 32). Another possible reason they failed to perform well comparatively to nonlinear measures is that the sources of electrophysiological signals (cells that generate electrical signal) are not independent; they are coupled, and that simple fact adds another level of complexity to the whole analysis and also to the interpretation of results.

This elaboration on the demonstrated performances of both standard and nonlinear measures in the task of detecting the CVD risks in depressive patients begs the answer to the questions why these effective research methods are not used in clinical practice and how much longer do we have to wait? One understandable reason for this delay in acceptance is that standard (or conventional) measures are already embedded in every single piece of equipment serving in electrophysiology, while fractal and other nonlinear measures are not, despite their very fast and low-cost calculation and interpretation. Another reason could be that the clinicians are not familiar with those measures, as they do not read the literature coming from technical sciences and still are of opinion that it is solely reserved for research.

To conclude, although Porges states (58) that psychiatrists and psychologists seem not to be sufficiently interested in the use of objective biomarkers in their daily diagnostic work with patients suffering from depression, the real question here is how ethical is it to keep this status quo and apply trial and error protocols in prescribing medications without prior objective screening for CVD risks?

**Conclusion**

The time has come to start using nonlinear methods of analysis of ECG since they have been repeatedly demonstrated to be superior over standard methods in early detection of the aberrated dynamics of coronary system. When the two most frequent causes of death and disability (cardiovascular diseases and depression) combine, the chances of survival significantly dicrease. These analytical methods are already in use in cardiology and provide good results in reducing this risk in cardiac patients that developed depression. It seems obvious that there is a need for the corresponding cardiologic screening in psychiatric patients.

**Acknowledgements**

Part of this work has been supported (MC) by the project RISEWISE (H2020-MSCA-RISE-2015-690874) and by the Ministry of Education, Science and Technological Development of the Republic of Serbia (DS).

**Conflict of Interest**: Authors have no conflict of interest to declare.